\documentclass[%
 reprint,
 superscriptaddress,
 amsmath,amssymb,
 aps,
 prd,
]{revtex4-1}

\usepackage{graphicx}
\usepackage{dcolumn}
\usepackage{bm}

\begin{document}

\preprint{APS/123-QED}

\title{Lateral Distribution of Muons in IceCube Cosmic Ray Events}
\affiliation{III. Physikalisches Institut, RWTH Aachen University, D-52056 Aachen, Germany}
\affiliation{School of Chemistry \& Physics, University of Adelaide, Adelaide SA, 5005 Australia}
\affiliation{Dept.~of Physics and Astronomy, University of Alaska Anchorage, 3211 Providence Dr., Anchorage, AK 99508, USA}
\affiliation{CTSPS, Clark-Atlanta University, Atlanta, GA 30314, USA}
\affiliation{School of Physics and Center for Relativistic Astrophysics, Georgia Institute of Technology, Atlanta, GA 30332, USA}
\affiliation{Dept.~of Physics, Southern University, Baton Rouge, LA 70813, USA}
\affiliation{Dept.~of Physics, University of California, Berkeley, CA 94720, USA}
\affiliation{Lawrence Berkeley National Laboratory, Berkeley, CA 94720, USA}
\affiliation{Institut f\"ur Physik, Humboldt-Universit\"at zu Berlin, D-12489 Berlin, Germany}
\affiliation{Fakult\"at f\"ur Physik \& Astronomie, Ruhr-Universit\"at Bochum, D-44780 Bochum, Germany}
\affiliation{Physikalisches Institut, Universit\"at Bonn, Nussallee 12, D-53115 Bonn, Germany}
\affiliation{Dept.~of Physics, University of the West Indies, Cave Hill Campus, Bridgetown BB11000, Barbados}
\affiliation{Universit\'e Libre de Bruxelles, Science Faculty CP230, B-1050 Brussels, Belgium}
\affiliation{Vrije Universiteit Brussel, Dienst ELEM, B-1050 Brussels, Belgium}
\affiliation{Dept.~of Physics, Chiba University, Chiba 263-8522, Japan}
\affiliation{Dept.~of Physics and Astronomy, University of Canterbury, Private Bag 4800, Christchurch, New Zealand}
\affiliation{Dept.~of Physics, University of Maryland, College Park, MD 20742, USA}
\affiliation{Dept.~of Physics and Center for Cosmology and Astro-Particle Physics, Ohio State University, Columbus, OH 43210, USA}
\affiliation{Dept.~of Astronomy, Ohio State University, Columbus, OH 43210, USA}
\affiliation{Dept.~of Physics, TU Dortmund University, D-44221 Dortmund, Germany}
\affiliation{Dept.~of Physics, University of Alberta, Edmonton, Alberta, Canada T6G 2G7}
\affiliation{D\'epartement de physique nucl\'eaire et corpusculaire, Universit\'e de Gen\`eve, CH-1211 Gen\`eve, Switzerland}
\affiliation{Dept.~of Physics and Astronomy, University of Gent, B-9000 Gent, Belgium}
\affiliation{Dept.~of Physics and Astronomy, University of California, Irvine, CA 92697, USA}
\affiliation{Laboratory for High Energy Physics, \'Ecole Polytechnique F\'ed\'erale, CH-1015 Lausanne, Switzerland}
\affiliation{Dept.~of Physics and Astronomy, University of Kansas, Lawrence, KS 66045, USA}
\affiliation{Dept.~of Astronomy, University of Wisconsin, Madison, WI 53706, USA}
\affiliation{Dept.~of Physics and Wisconsin IceCube Particle Astrophysics Center, University of Wisconsin, Madison, WI 53706, USA}
\affiliation{Institute of Physics, University of Mainz, Staudinger Weg 7, D-55099 Mainz, Germany}
\affiliation{Universit\'e de Mons, 7000 Mons, Belgium}
\affiliation{T.U. Munich, D-85748 Garching, Germany}
\affiliation{Bartol Research Institute and Department of Physics and Astronomy, University of Delaware, Newark, DE 19716, USA}
\affiliation{Dept.~of Physics, University of Oxford, 1 Keble Road, Oxford OX1 3NP, UK}
\affiliation{Dept.~of Physics, University of Wisconsin, River Falls, WI 54022, USA}
\affiliation{Oskar Klein Centre and Dept.~of Physics, Stockholm University, SE-10691 Stockholm, Sweden}
\affiliation{Department of Physics and Astronomy, Stony Brook University, Stony Brook, NY 11794-3800, USA}
\affiliation{Dept.~of Physics and Astronomy, University of Alabama, Tuscaloosa, AL 35487, USA}
\affiliation{Dept.~of Astronomy and Astrophysics, Pennsylvania State University, University Park, PA 16802, USA}
\affiliation{Dept.~of Physics, Pennsylvania State University, University Park, PA 16802, USA}
\affiliation{Dept.~of Physics and Astronomy, Uppsala University, Box 516, S-75120 Uppsala, Sweden}
\affiliation{Dept.~of Physics, University of Wuppertal, D-42119 Wuppertal, Germany}
\affiliation{DESY, D-15735 Zeuthen, Germany}

\author{R.~Abbasi}
\affiliation{Dept.~of Physics and Wisconsin IceCube Particle Astrophysics Center, University of Wisconsin, Madison, WI 53706, USA}
\author{Y.~Abdou}
\affiliation{Dept.~of Physics and Astronomy, University of Gent, B-9000 Gent, Belgium}
\author{M.~Ackermann}
\affiliation{DESY, D-15735 Zeuthen, Germany}
\author{J.~Adams}
\affiliation{Dept.~of Physics and Astronomy, University of Canterbury, Private Bag 4800, Christchurch, New Zealand}
\author{J.~A.~Aguilar}
\affiliation{D\'epartement de physique nucl\'eaire et corpusculaire, Universit\'e de Gen\`eve, CH-1211 Gen\`eve, Switzerland}
\author{M.~Ahlers}
\affiliation{Dept.~of Physics and Wisconsin IceCube Particle Astrophysics Center, University of Wisconsin, Madison, WI 53706, USA}
\author{D.~Altmann}
\affiliation{Institut f\"ur Physik, Humboldt-Universit\"at zu Berlin, D-12489 Berlin, Germany}
\author{K.~Andeen}
\affiliation{Dept.~of Physics and Wisconsin IceCube Particle Astrophysics Center, University of Wisconsin, Madison, WI 53706, USA}
\author{J.~Auffenberg}
\affiliation{Dept.~of Physics and Wisconsin IceCube Particle Astrophysics Center, University of Wisconsin, Madison, WI 53706, USA}
\author{X.~Bai}
\thanks{Physics Department, South Dakota School of Mines and Technology, Rapid City, SD 57701, USA}
\affiliation{Bartol Research Institute and Department of Physics and Astronomy, University of Delaware, Newark, DE 19716, USA}
% ...

\date{\today}
\collaboration{IceCube Collaboration}
\noaffiliation

\begin{abstract}
In cosmic ray air showers, the muon lateral separation from the center of the shower is a measure of the transverse momentum that the muon parent acquired in the cosmic ray interaction. IceCube has observed cosmic ray interactions that produce muons laterally separated by up to 400 m from the shower core, a factor of 6 larger distance than previous measurements.  These muons originate in high $p_T$ ($>$ 2 GeV/c) interactions from the incident cosmic ray, or high-energy secondary interactions. The separation distribution shows a transition to a power law at large values, indicating the presence of a hard $p_{T}$ component that can be described by perturbative quantum chromodynamics. However, the rates and the zenith angle distributions of these events are not well reproduced with the cosmic ray models tested here, even those that include charm interactions. This discrepancy may be explained by a larger fraction of kaons and charmed particles than is currently incorporated in the simulations.

\end{abstract}

\maketitle

\section{\label{sec:intro}Introduction}

There have been many attempts to measure the cosmic ray composition at energies around and above the knee of the spectrum ($\sim$ PeV) \cite{gaisser90,Bluemer:2009zf}.  At these energies, direct measurements by balloon and satellite experiments have very limited statistics.  Ground based experiments rely on indirect measurements using observables such as the ratio of the measured electromagnetic energy to the number of muons {\cite{IceCubecomposition,Antoni:2005wq}. These analyses are dependent on phenomenological calculations and simulations to relate the muon observations to an inferred composition; the result can be sensitive to the assumed hadronic interaction models \cite{Kang:2010bs}.

Studies of high-energy ($\gtrsim$ 1~TeV) muons with underground detectors have been an important part of this effort. The muons are produced early in air showers and probe the initial shower development \cite{Klein:2009ew}. Two classes of muons are generally considered. ``Conventional'' muons come from pion and kaon decays, while ``prompt'' muons come from the decays of particles containing heavy quarks, mostly charm.  Conventional muons dominate at TeV energies, but, at energies above 100 TeV, prompt muons are expected to dominate \cite{pasquali98}.  The resulting change in the slope of the muon energy spectrum has not yet been observed \cite{Icecubenumu11}. 

Studies of isolated muons, far from the shower core, can help understand the uncertainties due to phenomenological models.  Muon separations greater than about 30~m are largely due to the transverse momentum, $p_T$, imparted to the muon by its parent. For $p_T \gtrsim 2$ GeV/c, these interactions can be described in the context of perturbative quantum chromodynamics (pQCD). Data from RHIC, the Tevatron, and the LHC are in quite good agreement with modern fixed order plus next-to-leading log calculations \cite{vogt06}. These experimental studies give us some confidence in pQCD calculations for air showers. 

Experimentally, the transition from soft interactions ({\it i.e.} those with $p_{T} < 2$ GeV/c that are not describable in pQCD) to hard interactions is visible as a transition from a $p_T$ spectrum that falls off exponentially to one that follows a power law. At low $p_T$ the spectrum follows $\exp{(-p_T/T)}$, with $T\approx 220$ MeV/c for pions (somewhat higher for kaons and protons) \cite{Adare:2011vy}. At higher $p_T$, the spectrum falls as $1/(1+p_T/p_0)^n$, where one fit found n=$13.0^{+1.0}_{-0.5}$ and $p_0 =1.9 ^{+0.2}_{-0.1}$ GeV/c \cite{Adams:2004zg}. The transition is around 2 GeV/c for pions.  This spectral change should be visible in the lateral separation distribution.

The MACRO detector has previously measured the lateral separation between muons in air showers for primary energies ranging roughly from 10$^4$~GeV to 10$^6$~GeV \cite{ambrosio99}. Buried under 3.8 km water equivalent of rock, MACRO has a minimum muon energy of about 1.3 TeV. MACRO measured muon pair separations out to a distance of about 65~m. Their simulations verified the linear relationship between $p_T$ and separation (with a small offset due to multiple scattering of the muons) out to a $p_{T}$ of 1.2~GeV/c, below the expected transition to the pQCD regime. 

The muon $p_T$ is  related to the separation of the muon from the shower core by
\begin{equation}
d_{T}\eqsim\frac{p_{T}Hc}{E_{\mu}\cos(\theta)}
\label{eq:pt}
\end{equation}
where $d_T$ is the perpendicular separation between the muon and the shower core, $H$ is the interaction height of the primary, $H/\cos(\theta)$ is the path length of the shower to the ground at a zenith angle $\theta$, and $E_{\mu}$ is the energy of the muon at generation. The interaction height of the parent of the muon is assumed to be synonymous with the primary interaction height, and the energy of the muon at generation is well approximated by its energy at the surface of the Earth. 

In addition to the initial $p_T$, muons can separate from the shower core when they bend in the Earth's magnetic field or multiple scatter in the ice above the detector.  However, for the muon energies and separations considered here the gyroradius is on the order of 20,000 km and multiple scattering is similarly negligible; the initial $p_T$ is the dominant effect producing the separation. 

A selection of muons with large transverse separation is biased toward events produced at high altitudes, high $p_T$, and low energy. The detector geometry imposes a minimum energy; most of the muons will naturally be near this threshold.  In the competition between altitude and $p_T$, the atmospheric density decreases exponentially with increasing altitude, while the $p_T$ should fall more slowly, only as a power law.  Although widely separated muons are biased toward high-altitude interactions, they still allow for studies of transverse momentum.  Events with large zenith angles are also preferred, since the muons are given more time to separate.  Of course, all of these factors should be appropriately modeled in the Monte Carlo. 

In the IceCube neutrino telescope \cite{Halzen:2010yj}, muons are detected with a 1 km$^3$ array of optical sensors buried in the Antarctic ice at depths between 1450 and 2450~m; individual muons studied in this analysis must have an energy at the Earth's surface of at least 400~GeV to reach the detector. The 125~m horizontal spacing between IceCube strings serves as a rough threshold for the minimum resolvable separation. For vertical muons with an interaction height of 50~km and an energy of 1~TeV, this corresponds to a $p_{T}$ of 2.5~GeV/c. The interaction height and muon energy vary from shower to shower, so the event-by-event uncertainty in $p_{T}$ approaches a factor of 2 if we assume average values for both. 

The muon energy and cosmic ray interaction height for air showers that produce muons in the detector depend on the zenith angle. Figure \ref{fig:intht} shows the interaction height as a function of true zenith angle for showers at sea level for DPMJET simulation (see Secs. \ref{sec:sim} and \ref{sec:disc} for a full description of the simulation). This dependence arises in part because the column depth of the atmosphere and the primary energy for air showers increases with zenith angle, leading to higher interaction heights for horizontal showers. The majority of showers have muons contained within 135~m of the shower core, but a small fraction have muons with larger lateral extensions; these showers interact much higher in the atmosphere. Figure \ref{fig:minenergy} shows a fit to both the minimum and average energy of muons at the surface of the earth as a function of zenith angle, as well as the minimum energy calculated assuming continuous energy loss along the track ($dE/dx$). The energies shown are for the muon with the largest perpendicular distance from the shower core.

\begin{figure} [htb]
\includegraphics[width=0.48\textwidth]{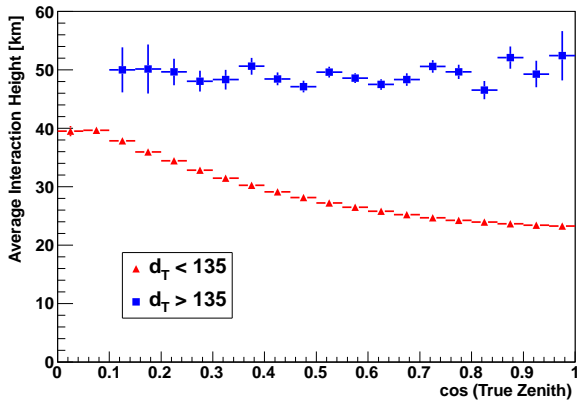}
\caption{\label{fig:intht} (Color online). The interaction height for all DPMJET simulated showers. Distributions for simulated showers with a true maximum muon separation less than 135~m and greater than 135~m are also shown.} 
\end{figure}

\begin{figure} [htb]
\includegraphics[width=0.48\textwidth]{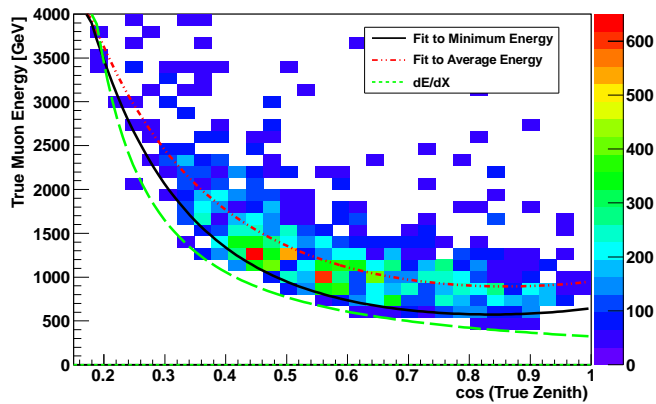}
\caption{\label{fig:minenergy} (Color online). The true energy at the surface of the Earth for the muon furthest from the shower core for simulated DPMJET shower events that pass all selection criteria versus zenith angle. Also shown are fits of the minimum and average energies. The values calculated using Eq. \ref{eq:emin} are also shown.}
\end{figure}

The zenith angle has an impact on shower development. The 1450~m of ice above the detector shield it from vertical muons with energies less than about 400~GeV. For inclined showers, several effects come into play.  The distance between the target and detector rises, giving the muon more time to separate from the shower core.  However, the slant depth also increases, raising the muon energy threshold roughly exponentially with the ice thickness.  For average $dE/dx$ energy loss, the minimum energy at the Earth's surface is given by \cite{chirkin04}
\begin{equation}
E ({\rm min}) = \frac{a}{b}\left[\exp(Db/\cos\theta) - 1\right]
\label{eq:emin}
\end{equation}
where $D$ is the depth of the detector and $a$ and $b$ are constants that describe the energy loss of a muon in ice (with values of 0.177 GeV/mwe and 0.209 $\times$ 10$^{-3}$/mwe, respectively) \cite{chirkin04}. 

The zenith angle distribution also depends on the parents of the muons.  At TeV energies, most muons originate from pions and kaons that decay before they interact. The probability of decay increases for larger zenith angles, because the pions and kaons spend more of their livetime at higher altitudes where the target density is lower so they are less likely to interact. The muon flux is \cite{gaisser90,desiati10}

\begin{align}
\frac{dN}{dE_\mu} \propto & \frac{0.14 E_{\mu}^{-2.7}}{\rm cm^{2}\,s\,sr\,GeV^{-3.7}} \bigg[\frac{1}{1+\frac{1.1E_\mu\cos(\theta)}{115 \rm GeV}} \nonumber\\
&+ \frac{0.054}{1+\frac{1.1E_\mu\cos(\theta)} {850 \rm GeV}} + \frac{9.1\times 10^{-6}}{1+\frac{1.0E_\mu\cos(\theta)} {5 \times 10^{7} \rm GeV}} \bigg]
\label{eq:angulardist}
\end{align}
where the three terms are for muons from pions, kaons, and charmed particles, respectively.  Charmed hadrons decay very quickly, leading to a flatter distribution in zenith angle.  This angular difference can be used to separate prompt muons from conventional muons; Fig. \ref{fig:pidzen} shows the zenith angle of cosmic ray muons with energies of $\sim$2~TeV produced by pion, kaon, or charm interactions. The fraction of muons from charm interactions increases for high zenith angles and higher energies.
\begin{figure} [htb]
\includegraphics[width=0.48\textwidth]{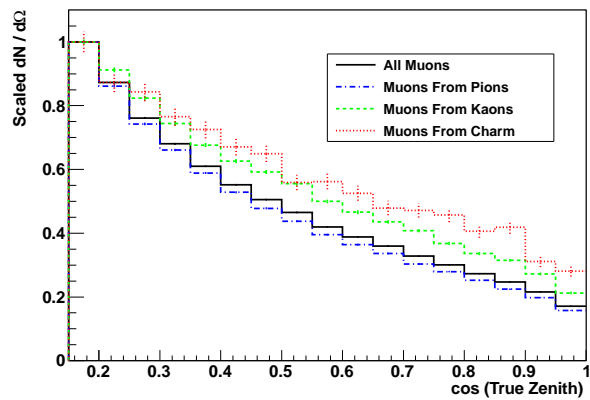}
\caption{\label{fig:pidzen} (Color online). The cosine of the zenith angle of DPMJET simulated muons produced by pion, kaon, or charmed particles interactions. The curves have been normalized to the peak bin and the muons energy is 2~TeV ($\pm 10\%$).}
\end{figure}

This paper extends the MACRO muon lateral separation measurements out to a separation of 400 m, well into the pQCD regime,  using 335 days of data collected with the partially completed IceCube detector.  The following sections give a description of the IceCube detector and an overview of the analysis.  The simulation is described in section \ref{sec:sim} and the background reduction is discussed in section \ref{sec:analysis}. The resulting distributions are discussed in sections \ref{sec:results} and \ref{sec:disc}.

\section{\label{sec:hardware}The IceCube Detector}

IceCube is a 1 km$^3$ underground neutrino telescope located at the South Pole.
5,160 digital optical modules (DOMs) on 86 vertical strings detect Cherenkov radiation from charged particles traversing the Antarctic ice \cite{Halzen:2010yj}. The DOMs are located between 1450 and 2450~m below the ice surface.  Each DOM consists of a 25~cm photomultiplier tube \cite{Abbasi:2010vc} plus associated digitization and calibration electronics \cite{Abbasi:2008ym}, all in a 35~cm diameter pressure vessel.

The primary energy and pseudorapidity of the muon with the maximum separation in simulated IceCube cosmic ray events that survive all selection criteria for this analysis are shown in Fig. \ref{fig:trueenergy}. The majority of muons studied here are produced in the far-forward region.

 \begin{figure} [htb]
\includegraphics[width=0.48\textwidth]{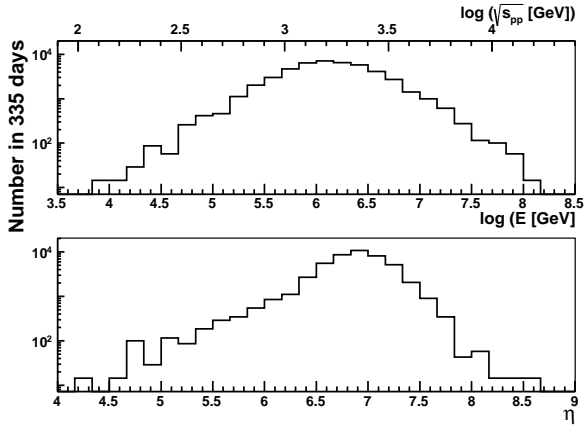}
\caption{\label{fig:trueenergy} Top: The true total primary energy of Sibyll simulated shower events that pass all selection criteria. The equivalent center of mass energy for proton-proton collisions is shown on the top axis. Bottom: The pseudorapidity of the muon with the maximum transverse separation in Sibyll simulated shower events that pass all selection criteria.}
\end{figure}

This analysis uses data taken from May 20, 2009 to May 30, 2010 when the array was partially complete, with 59 of the 86 strings deployed. Each DOM triggers internally at a threshold that gives it an 85\% efficiency for single photoelectrons; data is sent to the surface if the trigger occurs in coincidence with the DOM's nearest-neighbor or next-to-nearest-neighbor. There, a surface trigger selects events where 8 DOMs triggered within $10~\mu$s. The detector triggered at a rate of about 1600~Hz; the bulk of these events were from cosmic ray muons.  

\section{\label{sec:overview}Analysis Overview}

This analysis is sensitive to a shower with a bundle of low $p_T$ muons that are contained within a few 10s of meters of the shower core, plus an isolated high $p_T$ muon separated at least 125~m from the shower core.  The bundle gives a good reconstruction of the core location and primary zenith angle.  

 Each cosmic ray event is reconstructed with a two-track hypothesis. One track reconstructs the center of the muon bundle, and the other track reconstructs the muon with the largest lateral separation, which we call the laterally separated muon (LS muon). The perpendicular separation between the two tracks at the point of closest approach to the center of IceCube is defined as the lateral separation of muons in a cosmic ray shower ($d_T$ in Eq. \ref{eq:pt}). For showers without a distinct separation between the bundle core and the LS muon, this measurement can be thought of as the lateral extent of the cosmic ray air shower. Figure \ref{fig:muondist} shows the simulated true LS muon distance from the shower core at the initial filter level and after applying all selection criteria.  
\begin{figure} [htb]
\includegraphics[width=0.48\textwidth]{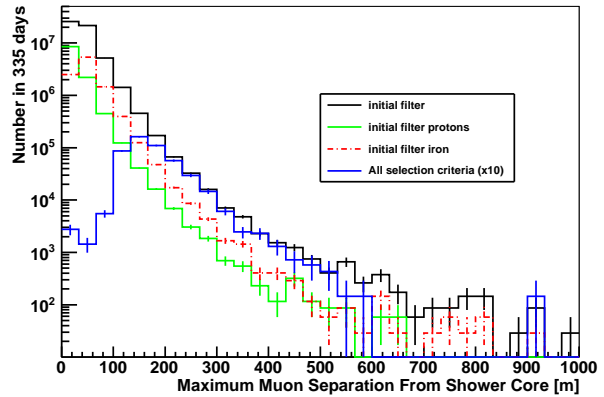}
\caption{\label{fig:muondist} (Color online). The Sibyll simulated true separation between the LS muon and the shower core for all cosmic ray showers, showers with proton primaries, and iron primaries at the initial filter level as well as the distribution after applying all selection criteria (described in Sec. \ref{sec:analysis}).}
\end{figure}
For showers with separations greater than $\sim$100~m, the spectral behavior of the iron and proton showers are very similar.   

  The background for this measurement is cosmic ray air showers without LS muons. There are two distinct types of background. The first is single showers from cosmic rays that appear as a single track in the detector and can be eliminated by requiring two high quality reconstructed tracks. Multiple independent coincident showers (``double showers'') constitute the second type of background. The IceCube 59-string configuration is large enough that the rate of simultaneous cosmic rays events is significant. Muon bundles from two (or more) uncorrelated air showers can deposit light in the detector within the event window, producing two (or multiple) separated tracks. The rate of multiple showers follows a Poissonian distribution, such that double cosmic ray events are the only significant background. The number of double showers can be reduced by requiring that the two reconstructed tracks arrive at roughly the same time (i.e. within $\pm$450~ns of each other) and from the same direction. However, an irreducible background remains from showers that arrive simultaneously from the same direction. This number is very small compared to the signal and can be measured by studying the data in the ``off-time'' window. Off-time events have tracks that arrive between 450~ns and 1350~ns of each other, after applying all other selection criteria.

\section{\label{sec:sim}Shower Simulation}
Cosmic ray air showers are generated using the CORSIKA \cite{heck98} simulation program. The simulated cosmic ray spectrum was based on the H\"{o}randel polygonato model \cite{horandel04} with energies between 600~GeV and $10^{11}$~GeV. The showers are generated using the Sibyll 2.1 \cite{ahn09}, DPMJET 2.55 \cite{ranft95}, and QGSJET01c \cite{kalmykov97} hadronic interaction models that are used to develop the shower in the atmosphere. QGSJET and DPMJET were chosen because they are currently the only models that include charm interactions. The current version of Sibyll does not include charm, but it is a modern model that reproduces relevant accelerator variables well, and is the default IceCube interaction model (see \cite{fedynitch12} for a detailed comparison of these models). The muons were propagated through the ice by the Muon Monte Carlo simulation package \cite{chirkin04}. The propagation of light through the ice and the response of the electronics, as well as the trigger, were simulated with IceCube software. All simulations were done with the CORSIKA MSIS-90-E atmospheres paramaterized for South Pole atmospheres measured in March, July, October, and December of 1997 \cite{nasatmo}.

Double shower events are simulated by combining single shower events with a time separation randomly drawn from an exponential distribution. The detector response to the double showers is simulated as one event. Events are counted as double showers even if they do not produce enough hits in the detector to reconstruct a track. This approach is conservative and results in an overestimation of the double shower rate for this analysis. This is acceptable because the double shower rate is measured using the data and the simulation serves only as a guide for event identification and development of selection criteria, and to calculate event rates at early cut levels before the two track reconstruction is performed.

Signal and background distributions are drawn from the same dataset, where showers with muons more than 100~m from the shower core that reach IceCube depths are considered LS muon signal. Roughly 23 days of simulated livetime have been generated with Sibyll, 11 days with QGSJET, and 7 days with DPMJET.

The majority of the figures shown in this paper are made with Sibyll because it has the largest simulated livetime. For most of the figures there is no significant difference between the different hadronic interaction models, so the lines for the other models are omitted for clarity. Distributions that required detailed knowledge of muon production inside the air shower were drawn from dedicated simulation that recorded the full interaction history of all muons. Due to computational limitations only the two hadronic models that included charm interactions, QGSJET and DPMJET, were used for the dedicated simulation.

A test simulation which varied ice absorption and scattering properties showed a 20\% increase in simulated event rate. However, this difference is a global shift, with no dependence on the variables shown in this paper, and is smaller than the variation in event rate for different hadronic interaction models, so this systematic is not included. Generating air showers is computationally intensive and computational resources are the primary limit on the livetime of generated simulation. Nonetheless, for this analysis the statistics generated are more than sufficient for comparison to data.

\section{\label{sec:analysis}Analysis}
This analysis uses the extremely high energy (EHE) filter stream where events are required to generate at least 630 photoelectrons \cite{Abbasi:2011ji}; the filter output rate is about 1.4 Hz. The photoelectron requirement corresponds to a primary energy threshold of about 1000~GeV, nearly an order of magnitude below the minimum energy of events that survive all the selection criteria (Fig. \ref{fig:trueenergy}) and is sufficiently low to avoid bias in the final selection. The EHE stream is sensitive to events from all directions and allows comparison of event rates as a function of zenith angle. After the EHE filtering, $6.4 \times 10^7$ events were left in the data sample. Additional reconstructions were run on the remaining events.

\subsection{\label{sec:reco}Reconstructing LS Muon Events}

LS muon events have a unique dual topology: a bundle of low $p_T$ muons that make up the core of the shower, and a laterally separated muon with the same direction and timing. These two components are reconstructed separately.

An initial, fast reconstruction based on a linear relationship between the arrival times and light wavefront gives the approximate direction and location of the bundle track. The hits are rotated into a plane perpendicular to the first guess track. The rotated hits are sorted into two sets using the k-means clustering algorithm, an algorithm that sorts the hits into two clusters according to their closeness to the mean of the cluster \cite{mackay03}. The larger set of hits is assumed to be from the muon bundle and is reconstructed with a maximum-likelihood function that accounts for the arrival time of the Cherenkov photons and the scattering of light in the ice \cite{ahrens04}. 

Hits that belong to the LS muon can be identified by their timing relative to the reconstructed bundle track. Since the LS muon arrives in the detector at roughly the same time as the bundle but at least a hundred meters from the shower core, its light has a much earlier arrival time than light could propagate from the shower core. Hits are considered LS muon hits if their arrival time is more than 100 ns earlier than the expectation for light from the bundle track. Additionally, LS muon hits are also required to be more than 90 m from the bundle fit to reduce the contamination of hits belonging to the bundle. The values for the timing and separation were chosen to minimize miscategorization of hits and maximize the number of events where two tracks can successfully be reconstructed.

Next, the process is iterated to increase the accuracy of the reconstruction. First, the hits are resorted according to their perpendicular distance from the reconstructed LS muon track. Any hits that are more than 100 m from the LS muon track are used to reconstruct a new bundle track. LS muon hits are selected based on the same timing and separation values used previously but relative to the new bundle track. For most events, the new reconstructed tracks are not very different, but iterating helps eliminate the few cases where the initial clustering algorithm did not perform well. 

Figure~\ref{fig:angleres} shows the cumulative fraction of events as a function of the space angle between the reconstructed and true track for Sibyll simulation. The reconstruction algorithms are able to reconstruct the direction of the bundle to within 4.0$^\circ$ and the LS muon to within 5.6$^\circ$ for 68\% of the events. Figure~\ref{fig:distres} shows the resolution of the measured separation between the two tracks for Sibyll simulation. The muon with the largest separation from the bundle center is used for the true value of the LS muon. The separation is measured with a resolution of $\sim$30~m. 

\begin{figure} [htb]
\includegraphics[width=0.48\textwidth]{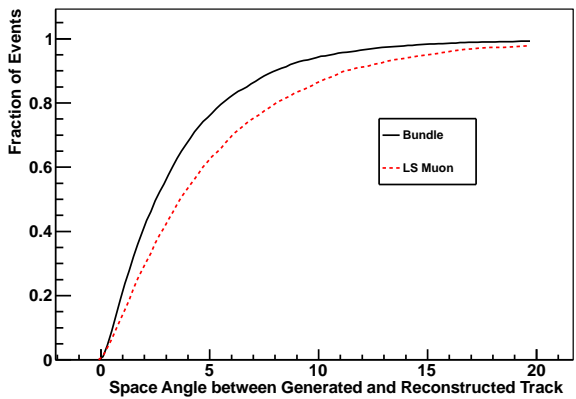}
\caption{\label{fig:angleres} (Color online). The angular resolution of the bundle and LS muon reconstructions for Sibyll simulation at the final selection level.}
\end{figure}

\begin{figure} [htb]
\includegraphics[width=0.48\textwidth]{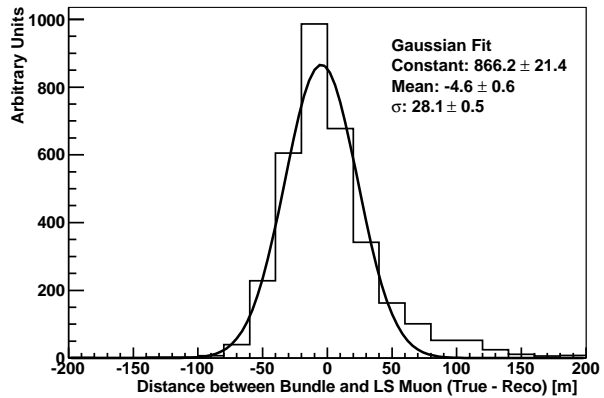}
\caption{\label{fig:distres} The separation resolution of the bundle and LS muon reconstructions for Sibyll simulation at the final selection level.}
\end{figure}

\subsection{Background Reduction}
Background reduction was studied with Sibyll since it has the largest livetime. Table \ref{tab:rates} shows the number of events passing each selection level.

After initial filtering, only events where the bundle and LS muon track reconstructions both succeeded are retained. This reduced the background from single showers by roughly two orders of magnitude while retaining more than 300,000 of the signal LS muon events. The main reason signal events fail to reconstruct is there are not enough hits in the LS muon track; the average number of hit DOMs from the LS muon is 20 (out of $\sim$3,500 DOMs on 59 strings). 

\begin{table*}
\caption{\label{tab:rates}Number of events passing each selection level in 335 days for data, the sum of all signal and background ($\Sigma$All), LS muons ($d_T > 100$), single showers and double showers estimated from Sibyll simulation (sim), and double showers estimated from off-time data. The uncertainties shown are statistical.}
\begin{ruledtabular}
\begin{tabular}{lrccccr}
\textrm{Selection Criterion}&
\textrm{Data}&
\textrm{$\Sigma$All}&
\textrm{LS Muons}&
\textrm{Single}&
\textrm{Double (sim)}&
\textrm{Double}\\
\colrule
EHE Filter & 63,649,372 & 5.47 $\pm$ 0.02 $\times$ $10^7$ & 1.77 $\pm$ 0.03 $\times$ $10^6$ & 4.31 $\pm$ 0.02 $\times$ $10^7$ & 9.79 $\pm$ 0.08 $\times$ $10^6$& - \\
Reco. Successful & 3,804,388 & 2.87 $\pm$ 0.01 $\times$ $10^6$ & 3.24 $\pm$ 0.02 $\times$ $10^5$ & 3.94 $\pm$ 0.02 $\times$ $10^5$ & 2.15 $\pm$ 0.01 $\times$ $10^6$ & -\\
$|{\Delta}T| < 450$ ns & 723,592 & 6.99 $\pm$ 0.03 $\times$ $10^5$ & 3.12 $\pm$ 0.02 $\times$ $10^5$ & 2.66 $\pm$ 0.02 $\times$ $10^5$ & - & 120,384\\
${\Delta}{\Phi} < 5^{\circ}$ & 327,962 & 3.24 $\pm$ 0.02 $\times$ $10^5$ & 1.84 $\pm$ 0.02 $\times$ $10^5$ & 1.38 $\pm$ 0.01 $\times$ $10^5$ & - & 1,445\\
Single Likelihood $>$ 7.5 & 69,877 & 7.54 $\pm$ 0.10 $\times$ 10$^4$ & 7.02 $\pm$ 0.10 $\times$ $10^4$ & 4.14 $\pm$ 0.24 $\times$ 10$^3$ & - & 1,020\\
$d_T$ $>$ 135 m  & 56,463 & 5.94 $\pm$ 0.09 $\times$ 10$^4$ & 5.74 $\pm$ 0.10 $\times$ $10^4$ & 1.08 $\pm$ 0.13 $\times$ $10^3$ & - & 905\\
LS Muon nDOMs $>$ 8 & 38,966 & 4.71 $\pm$ 0.08 $\times$ $10^4$ & 4.57 $\pm$ 0.08 $\times$ $10^4$ & 6.35 $\pm$ 0.96 $\times$ $10^2$ & - & 695\\
LS Muon nStrings $>$ 2 & 34,754 & 4.52 $\pm$ 0.08 $\times$ $10^4$ & 4.42 $\pm$ 0.08 $\times$ $10^4$ & 5.62 $\pm$ 0.90 $\times$ $10^2$ & - & 456\\
\end{tabular}
\end{ruledtabular}
\end{table*}

\subsubsection{Double Showers}
The background from double showers is reduced by requiring that the direction of the two reconstructed tracks agree to within 5$^{\circ}$ and that they arrive at the point of closest approach to the detector center within $\pm$450 ns of each other. An irreducible background remains from double showers that happen to come from the same direction at the same time. Requiring the two tracks to originate from the same direction and arrive at the same time reduced the rate of double showers by a factor of 1500, while reducing the simulated signal by a factor of two.
\subsubsection{Single Showers}
Elimination of single showers relies on the topological difference between single and double tracks. Single tracks are well-reconstructed by the likelihood functions made for a single track hypothesis while events with LS muons are not. Figure \ref{fig:like} shows the log of the output of the reduced likelihood function for a single track reconstruction (`single likelihood'), done using all the hits (bundle and LS muon), for single shower background, LS muon signal, and data. Events that are good fits to the single track reconstruction hypothesis have a lower value on this plot. Requiring a single likelihood greater than 7.5 reduced the background by a factor of 30 and retains 40\% of the signal.
\begin{figure} [htb]
\includegraphics[width=0.48\textwidth]{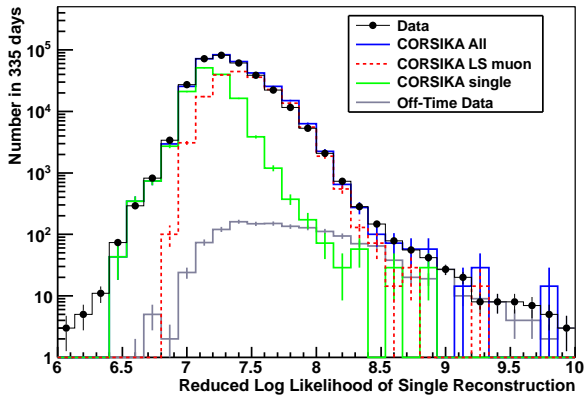}
\caption{\label{fig:like} (Color online). The reduced log likelihood for reconstruction with a single track hypothesis for data, all simulated Sibyll showers, LS muons, single showers, and double showers estimated from off-time data.}
\end{figure}

Next, events with a $d_T$ less than 135~m are removed. The closer the two tracks are, the more difficult it is for the reconstruction algorithm to separate them. Removing these closer tracks decreased the single shower background by a factor of four, while only removing 20\% of signal events; significantly improving the signal to background ratio.
\subsubsection{Angular Resolution}
After applying selection criteria to remove single and double showers, the predicted number of background events was about 1/30th the number of LS muon events. This was more than sufficient for measuring the LS muon separation spectrum, but some of the LS muon events still had poorly reconstructed directions. These LS muon tracks mostly triggered DOMs on a single string, making it difficult to reconstruct azimuth information. These events were eliminated by requiring that the LS muon hits occur on at least 3 IceCube strings. Also, the LS muon track was required to be robust, triggering more than 8 DOMs. These cuts select high quality events improving the angular resolution for the LS muon track from 7.3$^{\circ}$ to 5.6$^{\circ}$ while only removing 20\% of the signal events.

\section{\label{sec:results}Results}

After applying all the selection criteria, 34,754 events remain in data. Of those, the expected number of random double showers (based on off-time data rates) is 456. The number of predicted events depends on the interaction model. The Sibyll simulation predicts 44,800 $\pm$ 800 events, 98\% of which are LS muon events, while simulations with QGSJET and DPMJET predict 57,700 $\pm$ 1300 and 28,500 $\pm$ 1000 events, respectively (the uncertainties are statistical).

Figure \ref{fig:dist} shows the lateral distribution of data events that pass all selection criteria. The expectation from simulated LS muons is also shown and is in good agreement with the data. Additionally, the separation of double showers for off-time data events is shown. The separation distribution of double showers is flat, consistent with random coincidences (modulo detector edge effects). A linear fit to the ratios of simulation to data did not find a statistically significant difference in the slopes for the three simulation distributions, indicating they are all a reasonably good match for the data.
\begin{figure} [htb]
\includegraphics[width=0.48\textwidth]{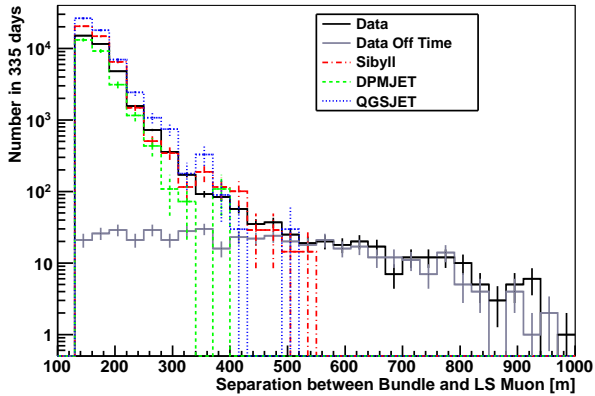}
\caption{\label{fig:dist} (Color online). The separation between the LS muon and bundle track after applying all selection criteria for data, simulation with the Sibyll, DPMJET, and QGSJET interaction models, as well as double showers estimated from off-time data.}
\end{figure}
Figure \ref{fig:zenslice} shows the lateral distribution of muons in cosmic ray showers for three different zenith slices after subtracting the separation distribution from the off-time events. The shape of the separation distribution does not show any strong dependence on zenith angle, indicating the separation is not a strong effect of propagation length.
\begin{figure} [htb]
\includegraphics[width=0.48\textwidth]{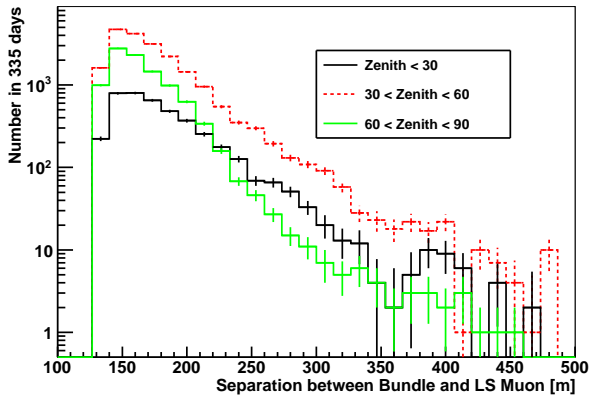}
\caption{\label{fig:zenslice} (Color online). The separation between the LS muon and bundle track for data events after applying all selection criteria and subtracting the off-time events for three zenith bands.}
\end{figure}

\subsection{Distribution at Sea Level}

The separation distribution shown in Fig. \ref{fig:dist} includes the effects of detector efficiency. This effect is quite large for two reasons. First, because of the steeply falling spectrum of cosmic rays, the majority of showers at the surface are too low in energy to generate muons that can reach IceCube depths. Second, there is a large geometrical effect from surface showers that do not intersect the detector. These combine to give correction factors that are as large as 10$^4$. However, the event-by-event variation is only about a factor of 10, primarily due to cut efficiency increasing with track separation.

To account for detector efficiency, we applied a bin-by-bin correction calculated from the ratio of the simulated separation distributions at sea level and after applying all selection criteria. To maximize statistics, and because no model was a perfect match for data, we used the unweighted average distribution from Sibyll, QGSJET, and DPMJET simulations. Statistical uncertainties were based on the number of events in each bin and were as high as 50\% of the correction value for bins with the lowest statistics. Systematic uncertainties were investigated by performing the same ratio for each model individually and for the distribution split into equal sized zenith bands. The systematic tests resulted in global normalization shifts, but the final distribution had the same shape and the same fit behavior described below, although with lower precision because of the reduced statistics.

The data shown in Fig. \ref{fig:dist} also includes a background of random double showers. This is removed by subtracting the distribution measured in off-time data (grey line) from the measured separation distribution (black line). The efficiency corrections are applied to the subtracted distribution and the result is shown in Fig. \ref{fig:scaled}.

\begin{figure} [htb]
\includegraphics[width=0.48\textwidth]{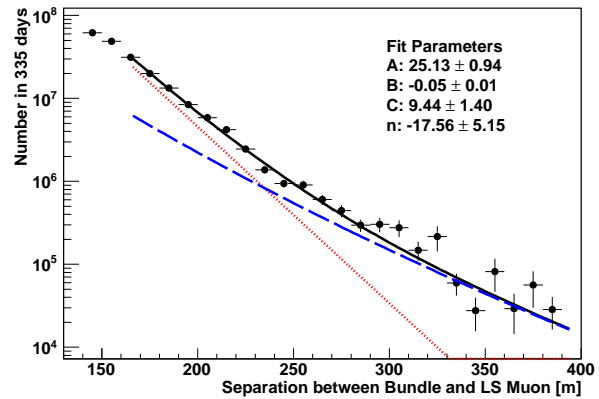}
\caption{\label{fig:scaled} (Color online). The LS muon data distribution at sea level, along with the best fit parameters for the function described in the text. The exponential part of the fit is plotted as a dotted red line and the power law is shown as a dashed blue line.}
\end{figure}

If the separation is proportional to $p_{T}$, the expected distribution would follow an exponential that transitions to a power law at large separations. To test this, the separation distribution was fit with this function:
\begin{equation}
  N=\exp(A+Bx)+10^{C}(1+x/400)^n
\end{equation}
with $A$, $B$, $C$, and $n$ allowed to vary. This function follows the form used in \cite{Adams:2004zg} to fit the $p_{T}$ distribution at RHIC. The resulting fit, shown in Fig. \ref{fig:scaled}, has a best fit with a transition at 235~m to a power law with an exponent of -17.6 $\pm$ 5.2. This composite fit has a ${\chi}^2$/DOF of 30.8/19, with a probability of 4\% of the model being a good fit for this distribution. While this value initially seems a bit low, values as low as 1\% are within the acceptable range \cite{press92}. A fit to a purely exponential function has a  ${\chi}^2$/DOF of 61.5/21 (probability 0.001\%).

For comparison, we have applied the same fit to the true separation at sea level for the Monte Carlo models.  For the sum of Sibyll, QGSJET, and DPMJET, the ${\chi}^2$/DOF was 17.0/19 for the two component fit, versus  ${\chi}^2$/DOF of 471/21 for an exponential fit.

\subsection{Zenith Angle Distribution}

A comparison of the arrival angle shows significant disagreement between simulation and data. Figure \ref{fig:zenith} shows the cosine of the reconstructed zenith angle of the bundle for events that survive all selection criteria. 
\begin{figure} [htb]
\includegraphics[width=0.48\textwidth]{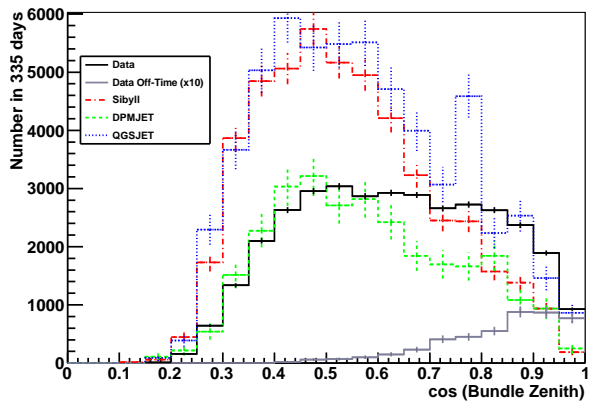}
\caption{\label{fig:zenith} (Color online). The cosine of the reconstructed bundle track after applying all selection criteria for data, simulated showers using the Sibyll, QGSJET, and DPMJET interaction models, as well as double showers estimated from off-time data (scaled by 10).}
\end{figure}
Sibyll and QGSJET overpredict the event rate at high zenith angles and underpredict the rate for the more vertical events, while DPMJET is a better match to the data at least at high zenith angles. There is no significant difference between the QGSJET and Sibyll distributions. Figure \ref{fig:zenratio} shows the ratio of simulation to data as a function of zenith angle. A linear fit to the ratio showed that DPMJET had the lowest slope, indicating it is the best match of the three simulations. A Kolmogorov-Smirnov (KS) test of the simulated distributions against the data finds that none of the simulations are a good match. The highest correlation is found for the DPMJET distribution, but the probability of it being drawn from the same distribution as the data was only $5 \times 10^{-12}$.

\begin{figure} [htb]
\includegraphics[width=0.48\textwidth]{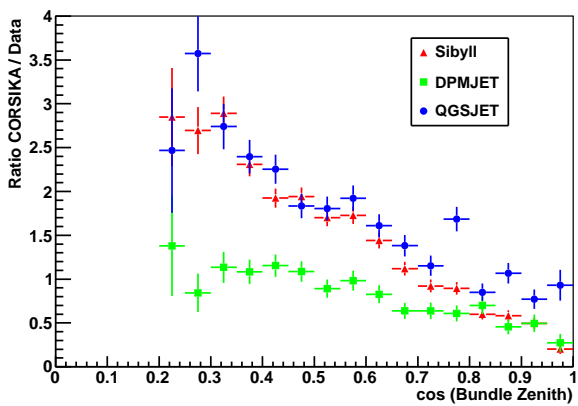}
\caption{\label{fig:zenratio} (Color online). The ratio of simulation to data versus the cosine of the zenith angle of the reconstructed bundle track after applying all selection criteria.}
\end{figure}

\section{Discussion} \label{sec:disc}
Figure \ref{fig:minpt} shows the minimum $p_T$ for a separation of 135~m calculated with the minimum and average muon energies (shown in Fig. \ref{fig:minenergy}) and the interaction height for showers with distant muons (from Fig. \ref{fig:intht}). 

DPMJET is a better match for the data at high zenith angles where the $p_{T}$ is expected to be slightly higher. However, even the DPMJET simulations underpredict the number of events at low zenith angles that also have high $p_{T}$ values. 
\begin{figure} [htb]
\includegraphics[width=0.48\textwidth]{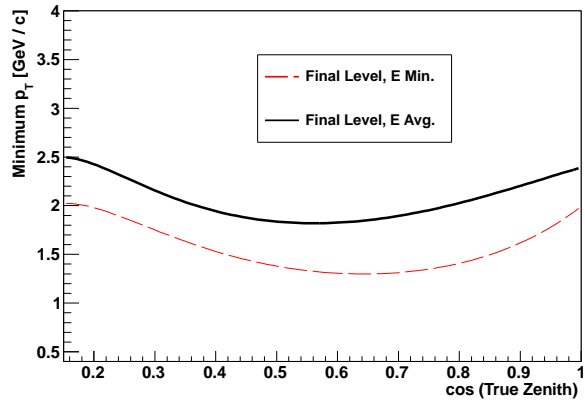}
\caption{\label{fig:minpt} (Color online). The minimum muon transverse momentum of DPMJET simulated shower events that pass all selection criteria for different energy parameterizations as a function of zenith angle. The interaction height comes from Fig. \ref{fig:intht}.}
\end{figure}

The different approaches to calculating $p_{T}$ taken by the various interaction models may contribute to the discrepancy in zenith angle. Sibyll 2.1 generates particle $p_T$ using two approaches \cite{ahn09}.  In soft (low $p_T$) interactions, quark-antiquark pairs are given a $p_T$ that follows a Gaussian distribution, with a mean of 0.3, 0.45, or 0.6 + $0.08\log_{10}{\sqrt{s/{\rm (30\ GeV)}}}$ GeV, for light quarks, strange quarks, and diquarks respectively; diquarks lead to the production of baryons.   Hard interactions are simulated using lowest order pQCD, with the GRV structure functions, including a saturation correction.  The minimum $p_T$ to qualify as a hard interaction increases as a function of energy. Charmed particles are not included in this version of Sibyll.  

DPMJET 2.55 also simulates both hard and soft interactions, with the hard interactions based on lowest order pQCD, using simple, phenomenological structure functions \cite{Berghaus:2007hp}. Charmed particles are produced both by soft processes, either at the ends of soft sea chains or inside the chain decay, and by ``hard processes at the ends of hard and semi hard chains (minijets)'' \cite{battistoni96}. The latter mechanism is modeled using pQCD, and the soft processes use phenomenological models. All possible charm particles are simulated.

QGSJET01c simulates interactions using cross sections calculated with an eikonal that is the sum of contributions from soft and hard processes \cite{kalmykov97}.  Charmed particles are produced when a charm quark-antiquark pair from the vacuum is coupled to the hadronizing strings of soft or semi-hard jets. Only the lightest charmed meson and baryon are generated (the $D$ and $\Lambda_c$). A comparison of predicted cross section and $p_T$ distributions with experimental measurements showed reasonable agreement at low $p_T$ \cite{goswami07}.

While all of these simulations include a hard component, it is interesting to note that DPMJET is the only model that includes a hard component of charmed particles, and it is also the model that agrees best with the data.

None of the current Monte Carlo codes include bottom quark production. At RHIC, bottom quark production is a significant contributor to the flux of leptons from heavy-flavor mesons at high $p_T$, contributing more than 20\% of the heavy lepton production with $p_T > 2$ GeV/c in proton-proton collisions at a center of mass energy of 200 GeV \cite{Aggarwal:2010xp}. Similar fractions are seen in 7~TeV proton-proton collisions at the LHC \cite{abelev12}. However, calculations of neutrino production from bottom interactions in cosmic ray showers yield rates around 2-3\% \cite{martin03}, in part because of the lower center of mass energy and smaller kinematic phase space in the far forward region. Muon production is expected to parallel the neutrino production so bottom quark production is only a small contributor to the overall muon flux.

The difference in zenith angle distribution may also indicate a larger fraction of muons from heavier parents. According to Eq. \ref{eq:angulardist}, the muon distribution becomes flatter as the parents become heavier; this behavior can be seen in Fig. \ref{fig:pidzen}. The flatter data distribution is consistent with a larger fraction of muons from kaons or charmed particles than is predicted in the simulation. A dedicated simulation that recorded the parents of all muons was performed for QGSJET and DPMJET. A full simulation of detector response was not possible, so events similar to events in the final sample were selected by requiring the muon energy to be greater than the minimum muon energy shown in Fig. \ref{fig:minenergy} and that the event have a muon at least 100~m from the shower core. Figure \ref{fig:kaonfrac} shows the fraction of muons with maximum separation from the bundle produced by pions, kaons, and charm particles in the shower. DPMJET has a higher fraction of muons produced by kaons at the zenith angles where the agreement with data is best. This suggests that muon production by kaons plays an important role in these events. Experimental measurements have shown a tendency for the kaon fraction to increase with target mass; various estimates of the kaon-pion ratio vary by as much as 20\% \cite{agrawal96}.

\begin{figure} [htb]
\includegraphics[width=0.48\textwidth]{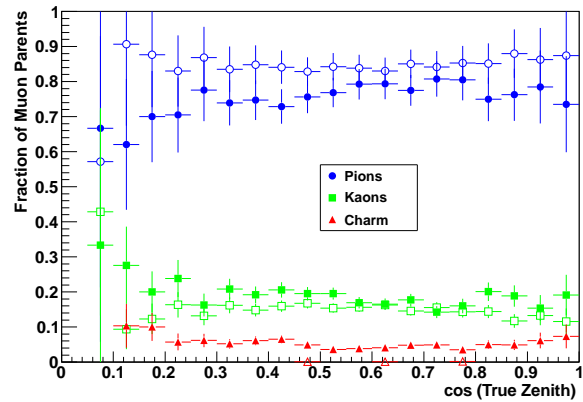}
\caption{\label{fig:kaonfrac} (Color online). The fractions of muons with maximum separation from the shower core from pions, kaons, and charm for DPMJET (closed shapes) and QGSJET (open shapes) for a sample of events similar to the final events in the data (see text for details).}
\end{figure}

The discrepancy in zenith angle may also be related to differences in the composition of the cosmic rays. Figures \ref{fig:comps} - \ref{fig:compd} show the ratios of the simulated cosine of the bundle zenith angle to data for Sibyll, QGSJET, and DPMJET for several different primary compositions. Sibyll and QGSJET show a flatter ratio for protons than for heavier elements. A KS test shows that the proton-only distributions are the best match to the data distribution for these two simulations. Contrastingly, the ratios with DPMJET simulations are all relatively flat with relatively similar KS probabilities. 

This suggests an interplay between kaon and charm abundance and composition. It is easier for proton primaries to produce high $p_{T}$ muons because all of their energy is concentrated in one particle. Sibyll and QGSJET, which have fewer kaons and (no) charmed particles, can only reproduce the data distributions with the cosmic rays most likely to produce high $p_{T}$ muons: protons. However, DPMJET simulation, with its greater fraction of kaons and charmed particles, is closer to the data distributions for all cosmic ray compositions.

\begin{figure} [htb]
\includegraphics[width=0.48\textwidth]{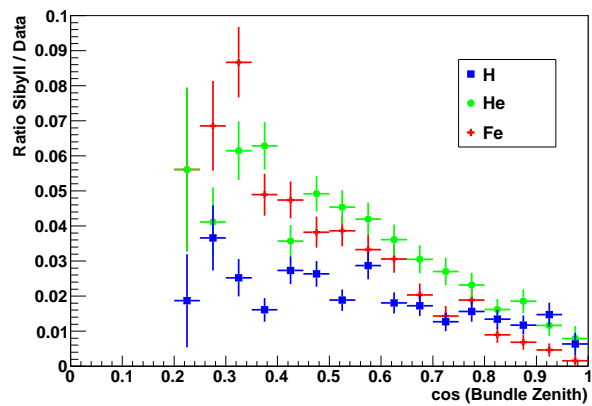}
\caption{\label{fig:comps} (Color online). The ratio of Sibyll simulation and data as a function of cosine of the reconstructed bundle zenith for different cosmic ray primary compositions.}
\end{figure}
\begin{figure} [htb]
\includegraphics[width=0.48\textwidth]{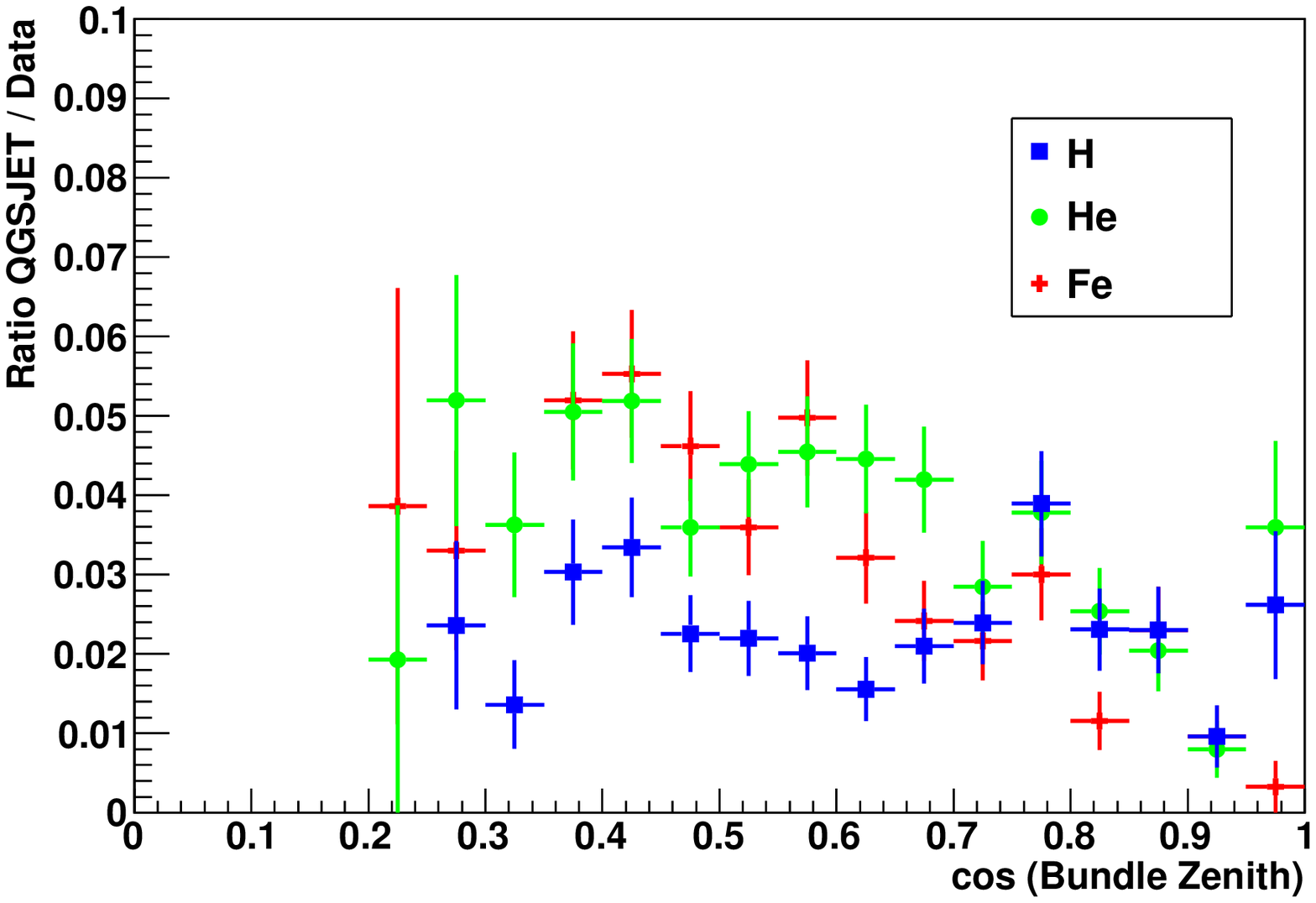}
\caption{\label{fig:compq} (Color online). The ratio of QGSJET simulation and data as a function of cosine of the reconstructed bundle zenith for different cosmic ray primary compositions.}
\end{figure}
\begin{figure} [htb]
\includegraphics[width=0.48\textwidth]{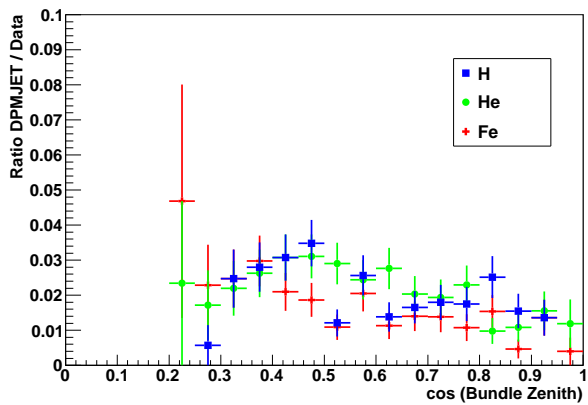}
\caption{\label{fig:compd} (Color online). The ratio of DPMJET simulation and data as a function of cosine of the reconstructed bundle zenith for different cosmic ray primary compositions.}
\end{figure}

The better agreement with data for protons simulated with Sibyll and QGSJET may indicate a dependence on the relative abundances of cosmic ray primaries. A cosmic ray composition model with lighter composition between 1 to 10 PeV could also improve the agreement in zenith angle for these models.

The zenith angle distribution may also depend on the inelasticity in ${\pi}N$ and $KN$ interactions. If pions and kaons lose less energy than predicted when they interact, then the high-energy muon flux may be enhanced, especially near the vertical direction.

\section{Conclusion}
IceCube has observed 34,754 muons with lateral separations between 135 m and 400 m; this corresponds to a transverse momentum of at least 2 GeV/c. The separation distribution is poorly fit by an exponential distribution with a ${\chi}^2$/DOF of 61.5/21.  The fit improves when a power law component is included to a ${\chi}^2$/DOF of 30.8/19, as expected from pQCD. However, the zenith angle distribution of the muons is unexpectedly flat, even when including the decay of charmed particles, and is poorly modeled by current simulations. This may be caused by an underproduction of kaons and charmed particles in the simulation. Future simulations with more sophisticated $p_{T}$ modeling may improve the disagreement in zenith angle.

IceCube has demonstrated the capability to resolve laterally separated muons in air showers. When improved simulation becomes available, future analyses could generate an estimate of muon parent ratios as well as a measurement of the transverse momentum spectrum in cosmic ray air showers.

\begin{acknowledgments}
We acknowledge the support from the following agencies:
U.S. National Science Foundation-Office of Polar Programs,
U.S. National Science Foundation-Physics Division,
University of Wisconsin Alumni Research Foundation,
the Grid Laboratory Of Wisconsin (GLOW) grid infrastructure at the University of Wisconsin - Madison, the Open Science Grid (OSG) grid infrastructure;
U.S. Department of Energy, and National Energy Research Scientific Computing Center,
the Louisiana Optical Network Initiative (LONI) grid computing resources;
National Science and Engineering Research Council of Canada;
Swedish Research Council,
Swedish Polar Research Secretariat,
Swedish National Infrastructure for Computing (SNIC),
and Knut and Alice Wallenberg Foundation, Sweden;
German Ministry for Education and Research (BMBF),
Deutsche Forschungsgemeinschaft (DFG),
Research Department of Plasmas with Complex Interactions (Bochum), Germany;
Fund for Scientific Research (FNRS-FWO),
FWO Odysseus programme,
Flanders Institute to encourage scientific and technological research in industry (IWT),
Belgian Federal Science Policy Office (Belspo);
University of Oxford, United Kingdom;
Marsden Fund, New Zealand;
Australian Research Council;
Japan Society for Promotion of Science (JSPS);
the Swiss National Science Foundation (SNSF), Switzerland.
\end{acknowledgments}

\bibliography{lsmuons}

\end{document}